\documentclass[
    aip,
    rsi,
    reprint,
    amsmath,
    amssymb,
    floatfix
]{revtex4-2}

\usepackage[utf8]{inputenc}
\usepackage[T1]{fontenc}
\usepackage{lmodern}
\usepackage{graphicx}
\usepackage{bm}
\usepackage{booktabs}
\usepackage{xcolor}
\usepackage{hyperref}
\usepackage{svg}
\hypersetup{hidelinks}
\begin{document}

\title{Time-Deterministic and Instruction-Dynamic Execution: A field-programmable gate array quantum-control processor}

\author{Xiaoqin Luo}
\affiliation{Beijing Academy of Quantum Information Sciences, Beijing 100193, China}

\author{Jiayun Song}
\affiliation{State Key Laboratory of Low Dimensional Quantum Physics, Tsinghua University, Beijing 100084, China}

\author{Xiaolu Su}
\email{suxiaolu0810@gmail.com}
\affiliation{Beijing Academy of Quantum Information Sciences, Beijing 100193, China}

\begin{abstract}
Long-running quantum computations, including error-corrected workloads, place joint demands on instruction availability, measurement-dependent control, and calibration updates. We present Time-Deterministic and Instruction-Dynamic Execution (TIDE), a field-programmable gate array (FPGA) quantum-control processor supporting runtime revision at both instruction and parameter levels. Dynamic Instruction Parameter Update (DIPU) supplies one-shot, next-match patches before parameter capture, while Dynamic Instruction Stream Overwrite (DISO) provides guarded replacement, logical deletion, and out-of-line insertion in eligible future regions of a resident program. A moving fetch-visibility guard coordinates instruction updates with execution, and event commitment isolates complete descriptors for independent hardware-timed dispatch. Revision applicability depends on update timing and execution progress; deterministic service applies to committed events under the stated scheduling conditions. The Xilinx ZCU102 implementation uses a 250-MHz control core and a 425-MHz timing/update domain. FPGA-resident measurements at the registered digital interfaces confirmed programmed-cycle dispatch for all tested descriptors committed at least one timing-domain cycle before their timestamps, with downstream ready and the stated first-in, first-out (FIFO) ordering conditions. Separate tests confirmed post-commit timestamp and payload invariance. The minimum all-success mapped DIPU margin was four control-domain cycles, and a replacement of \(L\) instruction words completed in \(L+5\) update-domain cycles under continuous payload delivery. Within the tested guard-distance range, rejected DISO requests reported an error and preserved the resident path, whereas all admitted revisions tested here executed complete revised sequences. These results provide a hardware execution basis for incorporating runtime-generated sequences and parameter corrections into precisely timed quantum computations with feedback.
\end{abstract}


\maketitle

\section{Introduction}

Executing long, adaptive quantum computations requires a sustained supply of control instructions and the timely application of classical decisions. Real-time quantum error correction (QEC), measurement-dependent logical operations, and calibration under drift bring these demands into the same execution layer~\cite{RyanAnderson2021RealTimeFTQEC,Caune2026RealTimeQEC,Google2025SurfaceCodeThreshold,Corcoles2021DynamicCircuits,Baumer2024LongRangeDynamicCircuits,Bluvstein2024LogicalProcessor,Berritta2025HamiltonianTracking,Sivak2026RLQEC}. Control electronics must incorporate new instructions and parameter values into future operations while preserving the schedule of events already committed for dispatch.

Control sequences with substantial nonrepeating content can exceed fast on-chip instruction storage, making the delivery of successive blocks part of execution~\cite{AntQ2026}. Measurement-dependent control can determine the content of later operation blocks as new results become available~\cite{RyanAnderson2021RealTimeFTQEC,Caune2026RealTimeQEC,Google2025SurfaceCodeThreshold,Corcoles2021DynamicCircuits,Baumer2024LongRangeDynamicCircuits,Bluvstein2024LogicalProcessor}. Supplying instruction blocks during a run allows the controller to incorporate these continuations alongside the resident program~\cite{Niroula2025QuantumStreaming}. Adaptive calibration likewise translates updated device estimates into control settings: Hamiltonian and relaxation-rate tracking adjust drive frequencies and interrogation settings~\cite{Berritta2025HamiltonianTracking,Berritta2026RelaxationTracking}, while QEC statistics can guide physical-control optimization~\cite{Sivak2026RLQEC}. These settings must take effect at the intended future operations.

\begin{figure*}[!t]
\centering
\includegraphics[width=\textwidth]{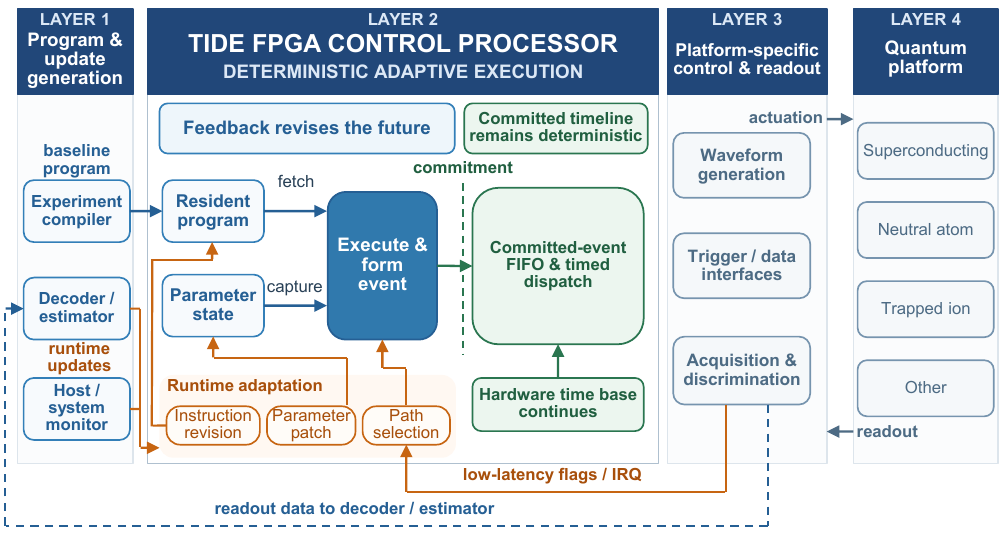}
\caption{TIDE in a conceptual quantum-control stack, connecting program and update sources to platform-specific electronics. Resident programs can be prepared with the prototype reference assembler; FPGA-resident traffic-generation logic supplied runtime updates during characterization. Future parameters and instructions remain revisable before their capture or fetch-visibility boundaries. Complete descriptors enter independent hardware-timed dispatch after commitment. External acquisition, decoding, waveform-generation, and quantum-platform paths indicate prospective integration. IRQ denotes interrupt request.}
\label{fig:system}
\end{figure*}

Field-programmable gate array (FPGA) and radio-frequency system-on-chip platforms such as Advanced Real-Time Infrastructure for Quantum physics (ARTIQ), the Quantum Instrumentation Control Kit (QICK), and QubiC provide hardware-timed control, readout, and feedback~\cite{Salathe2018LowLatency,kasprowicz2020artiq,xu2021qubic,stefanazzi2022qick,Yang2022FPGAControl}. Open Quantum Assembly Language (OpenQASM)~3, executable quantum instruction sets, and queue-based microarchitectures express classical control and timing semantics~\cite{Cross2022OpenQASM3,Fu2019eQASM,Fu2017QuMA}. Resident branches in the Arbitrary Pulse Sequencer~2 (APS2), the executable quantum instruction set architecture eQASM, and QubiC and conditional pulse gating in Presto implement measurement-dependent choices~\cite{Ryan2017DynamicQuantum,Fu2019eQASM,Xu2023QubiC2,Fruitwala2024DistributedQubiC,Tholen2022Presto}. Parameter stitching and Quantum Intermediate Representation (QIR)-based execution support numerical updates and runtime arithmetic~\cite{Rajagopala2024PCE,Lubinski2022RealtimeHybrid}. ARTERY uses prediction and pre-execution to reduce resident-path latency, while RISC-Q combines a RISC-V controller with timed first-in, first-out (FIFO) parameter queues and preloaded alternatives~\cite{Tian2025ARTERY,Liu2025RISCQ}. Timed queues decouple instruction processing from output timing~\cite{stefanazzi2022qick,Tholen2022Presto}; reconfigurable mappings, distributed synchronization, and vectorized control support larger systems~\cite{Fruitwala2024DistributedQubiC,Zhang2024ClassicalArchitecture,Liu2024FPGAScheduling,Silva2026Manarat,Zhao2025DistributedHISQ,VectorQCP2026}.

Runtime instruction and gate-data delivery is addressed through several architectures. Ant-Q combines a dynamic random-access memory (DRAM)--block random-access memory (BRAM) hierarchy with pipelined command delivery for long control sequences~\cite{AntQ2026}. Irtija et al.\ describe coordinated partial reprogramming of gate-sequencer lookup tables~\cite{Irtija2023SoCCommunication}, and the proposed NVQLink architecture accommodates dynamic instruction queues and operation streaming~\cite{caldwell2025platformarchitecturetightcoupling}. In the experiment of Niroula et al., the Helios runtime compiles gate lists from incoming external data into physical operations during execution~\cite{Niroula2025QuantumStreaming}. Table~\ref{tab:feedback_comparison} summarizes selected reported metrics and their measurement boundaries. For a running resident program, the execution challenge is to coordinate newly supplied content with advancing fetch and already committed events.

\begin{table*}[t]
\caption{Runtime-feedback capabilities and selected reported metrics of quantum-control architectures. The entries distinguish controller latency, instruction cost, event-submission interval, update-completion time, and required update lead time. Each metric retains the measurement boundary used in its source; the entries do not constitute a common end-to-end benchmark. TIDE's DISO completion times were measured under continuous payload delivery for the replacement lengths tested in Sec.~\ref{sec:diso_characterization}. RTIO denotes real-time input/output; DMA, direct memory access; TTL, transistor--transistor logic; CPU, central processing unit; ADC, analog-to-digital converter; and DAC, digital-to-analog converter. ACPKI names the cache-coherent-port kernel initiator of the cited Zynq implementation.}
\label{tab:feedback_comparison}
\centering

\begingroup
\small
\setlength{\tabcolsep}{3.5pt}
\renewcommand{\arraystretch}{1.10}

\begin{tabular}{
@{}
p{0.12\textwidth}
p{0.28\textwidth}
p{0.27\textwidth}
p{0.27\textwidth}
@{}
}
\toprule

\textbf{Platform}
&
\textbf{Runtime-feedback capability}
&
\textbf{Reported metric}
&
\textbf{Measurement boundary}
\\

\midrule

ARTIQ on Zynq~\cite{kasprowicz2020artiq,Lam2021ARTIQZynq}
&
Timestamped RTIO execution with input-dependent control and DMA playback.
&
376~ns sustained RTIO event-submission interval through ACPKI; 32~ns per TTL event during DMA playback.
&
ZC706: CPU-to-RTIO event submission and prerecorded-event playback, respectively; neither is an end-to-end feedback latency.
\\

\addlinespace[2pt]

eQASM~\cite{Fu2019eQASM}
&
Measurement-conditioned program flow and feedback control.
&
$\sim 92$~ns fast conditional execution; $\sim 316$~ns comprehensive feedback control.
&
Measurement result entering the Central Controller to the corresponding feedback-dependent digital output.
\\

\addlinespace[2pt]

QICK~\cite{stefanazzi2022qick}
&
Measurement-conditioned branching with hardware-timed pulse execution.
&
42~ns conditional-evaluation/address-jump component; 52~ns next-pulse component; 184--211~ns reported aggregate latency (2022 implementation).
&
Controller-side conditional execution; the aggregate latency additionally includes ADC/DAC contributions.
\\

\addlinespace[2pt]

QubiC~2.0~\cite{Xu2023QubiC2}
&
Measurement-dependent control flow and pulse-to-pulse parameter updates.
&
500~MHz processor; pulse commands require $\geq 4$ cycles and most register-write/control-flow instructions require 4--7 cycles.
&
Processor-core instruction execution; a directly corresponding end-to-end feedback latency is not reported.
\\

\addlinespace[2pt]

TIDE
&
DIPU parameter revision and guarded DISO replacement, deletion, and insertion.
&
Minimum all-success DIPU margin of four 250~MHz control cycles; DISO replacement of $L$ instruction words completes in ($L+5$) cycles at 425 MHz frequency.
&
Processor-facing runtime-update interface to parameter capture or instruction-update completion.
\\

\bottomrule
\end{tabular}

\endgroup
\end{table*}

We present Time-Deterministic and Instruction-Dynamic Execution (TIDE), an FPGA quantum-control processor that coordinates runtime program revision with parameter capture and event commitment (Fig.~\ref{fig:system}). Dynamic Instruction Parameter Update (DIPU) supplies a one-shot patch to the next matching event when the patch is available at parameter capture. Dynamic Instruction Stream Overwrite (DISO) revises eligible future program regions through replacement, logical deletion, and out-of-line insertion. Target-relative admission and a moving visibility guard coordinate revised instructions with the advancing fetch path, while the Time-Deterministic Peripheral-Control (TDPC) engine independently dispatches committed descriptors. This coordination keeps future control revisable while isolating committed timestamps and payloads from subsequent updates.

In the ZCU102 implementation, tests at the registered digital interfaces confirmed programmed-cycle dispatch under the specified scheduling and output conditions, together with post-commit timestamp and payload invariance. The minimum all-success mapped DIPU margin was four 250-MHz control-domain cycles; under continuous payload delivery, a replacement of \(L\) instruction words completed in \(L+5\) cycles at 425~MHz. Independent post-route sweeps characterized channel and program-memory capacity scaling (Table~\ref{tab:scaling}).

\section{Processor Architecture and Execution Contract}
TIDE defines three state boundaries. Parameter values remain revisable
until capture into an event descriptor; resident instructions remain
revisable while their target region can be protected from fetch; and
complete descriptors become immutable when accepted by a TDPC FIFO.
These parameter-capture, instruction-visibility, and event-commitment
boundaries define an execution contract that separates runtime revision
from hardware-timed dispatch.

\begin{figure*}[!t]
\centering
\includegraphics[width=0.98\textwidth]{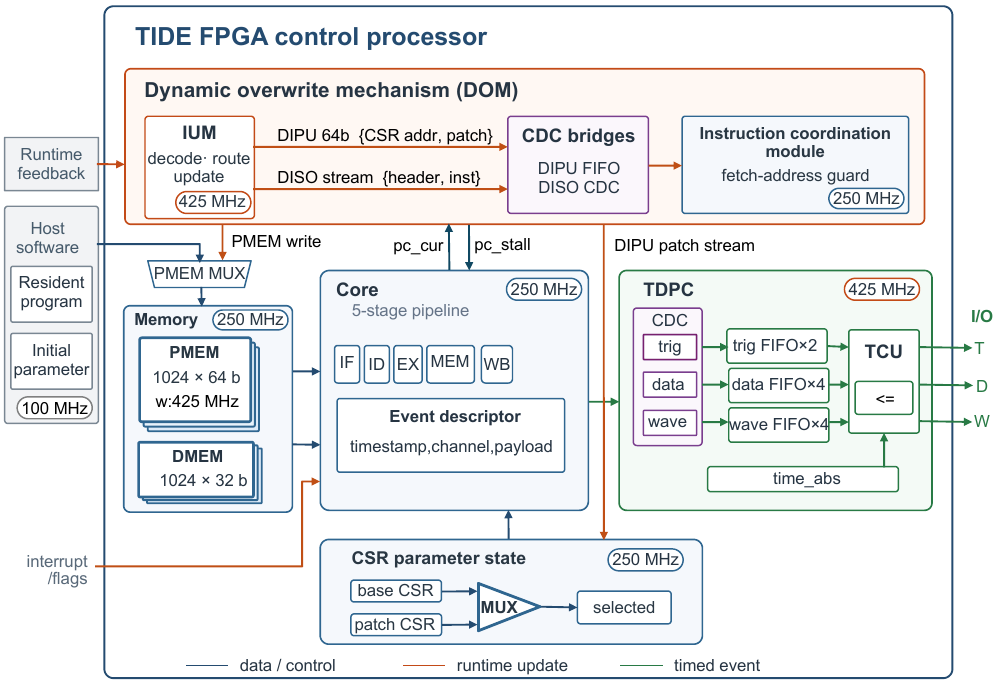}
\caption{Architecture and clock-domain organization of TIDE.
Runtime DIPU and DISO transactions enter the Instruction Update Module
(IUM), while the Instruction Coordination Module (ICM) guards
program-memory visibility and locally stalls fetch when required. The
250~MHz control core forms complete timestamped descriptors from
resident instructions and selected CSR values. Valid--ready acceptance
into a per-channel TDPC FIFO marks event commitment, after which the
425~MHz Timing Control Unit (TCU) dispatches registered trigger,
data-control, and waveform-parameter transactions against the
free-running \texttt{time\_abs} counter. The runtime-feedback arrow represents both independent processor-facing interfaces, DIPU and DISO, listed in Table~\ref{tab:processor_config}. Clock labels indicate the 100,
250, and 425~MHz domains. PMEM and DMEM denote program and data memory; CSR, control and status register; CDC, clock-domain crossing; MUX, multiplexer; and I/O, input/output. The pipeline labels IF, ID, EX, MEM, and WB denote instruction fetch, instruction decode, execution, memory access, and writeback, respectively. T, D, and W label trigger, data, and waveform outputs. The signals \texttt{pc\_cur}, \texttt{pc\_stall}, and \texttt{time\_abs} denote the current fetch address, local fetch-stall control, and hardware time count, respectively.}
\label{fig:system_overview}
\end{figure*}

Figure~\ref{fig:system_overview} shows the architecture. Before execution,
host software loads the resident program and initializes the base
parameter state. During execution, runtime feedback may supply parameter
or instruction updates, while condition flags and interrupts select
among resident control paths. The Dynamic Overwrite Mechanism (DOM)
applies these updates within the capture and visibility boundaries.

\subsection{Programmable Control Core and Event Formation}

The programmable control core executes the resident experiment program
upstream of event commitment. Its cache-free, five-stage in-order
pipeline preserves program order without cache-induced timing
variability.

Table~\ref{tab:instruction_classes} summarizes the supported instruction
classes.

A peripheral-oriented instruction does not directly actuate an
output. Instead, it specifies an event class, output channel, timing
information, and a parameter set bound to control and status registers (CSRs). The control-core
event-formation path reads the selected parameter values, combines
the instruction timing with the active temporal reference, and
assembles a complete descriptor containing the channel, effective
timestamp, and class-dependent payload. Capture of the selected
values into this descriptor defines the parameter-capture boundary.
After capture, the descriptor fields remain stable even if
back-pressure delays acceptance by the corresponding TDPC FIFO.
FIFO acceptance marks the separate event-commitment boundary
(Sec.~II\,B).

Binding event parameters through CSRs decouples the resident instruction
structure from operation-specific numerical values. Section~II\,C
specifies base-state writes, temporary patches, and their lifecycle and
same-target flow-control semantics.

\begin{table*}[t]
\caption{Functional instruction classes supported by the TIDE
control core.}
\label{tab:instruction_classes}
\centering
\begingroup
\small
\setlength{\tabcolsep}{4pt}
\renewcommand{\arraystretch}{1.18}
\begin{tabular}{%
@{}
p{0.16\textwidth}
p{0.28\textwidth}
p{0.50\textwidth}
@{}
}
\toprule
\textbf{Class}
&
\textbf{Representative instructions}
&
\textbf{Role in TIDE}
\\
\midrule

General-purpose
&
\texttt{add}, \texttt{mul}, \texttt{lw}, \texttt{sw},
\texttt{beq}, \texttt{jal}
&
Provides arithmetic, memory access, parameter calculation,
loops, branches, and resident experiment sequencing.
\\

Time-oriented
&
\texttt{sync}, \texttt{synci}, \texttt{syncr}, \texttt{wait},
\texttt{time\_abs}
&
Establishes or reads temporal references, observes the hardware
time base, and expresses absolute or stage-relative timing.
\\

Peripheral-oriented
&
\texttt{set\_trigger}, \texttt{set\_data}, \texttt{set\_wave}
&
Initiates the formation of timestamped trigger, data-control,
and waveform-parameter descriptors for TDPC commitment and
dispatch.
\\

CSR and feedback control
&
\texttt{csrr}, \texttt{csrw}, \texttt{csrwi},
\texttt{csrwbs}, \texttt{mret}
&
Accesses parameter and status state and supports
interrupt-driven resident control flow.
\\

\bottomrule
\end{tabular}
\endgroup
\end{table*}

\subsection{Committed-Event Timing and Dispatch}

The Time-Deterministic Peripheral-Control (TDPC) engine controls event
timing after commitment. It comprises independent per-channel FIFOs for
trigger-, data-control-, and waveform-parameter descriptors, together
with a Timing Control Unit (TCU) driven by the free-running 48-bit
hardware counter \texttt{time\_abs}. Successful valid--ready acceptance
of a complete descriptor by the selected TDPC FIFO marks event
commitment. Its timestamp and payload then remain unchanged by
subsequent control-core execution, parameter updates, or program-memory
revision. Control-core stalls and DOM fetch-protection activity do not
gate the hardware time base or committed-event dispatch path.

Each channel is scheduled from the head of its own FIFO. Trigger
descriptors carry digital-control fields, data descriptors carry
peripheral-control words, and waveform-parameter descriptors carry
parameter blocks for downstream waveform-generation modules.
All descriptor classes follow the same scheduling rule. The TCU compares
each FIFO-head timestamp with \texttt{time\_abs} and presents the
registered output transaction at the programmed issue cycle, provided
that the descriptor was committed with sufficient lead time. Dispatch at
the characterized processor boundary is defined by completion of the
output valid--ready handshake.

If a descriptor does not enter the TDPC FIFO with the required
scheduling margin, the present implementation retains it and
presents it at the first eligible output cycle rather than
discarding it. Cycle-exact dispatch therefore requires both timely
event commitment and downstream readiness at the programmed cycle.
A full TDPC FIFO back-pressures the descriptor-transfer path rather than
dropping events. The pending descriptor remains stable until acceptance.
This wait does not itself block formation of unrelated later descriptors,
subject to available upstream buffering. Because dispatch proceeds from
the FIFO head, descriptors must be supplied in nondecreasing timestamp
order within each channel. Independent channel FIFOs prevent occupancy
on one channel from blocking FIFO-head processing on another. A registered
channel completes at most one output handshake per timing-domain cycle.
Cycle-exact service therefore also requires at most one descriptor per
channel per timestamp and no older head-of-line descriptor retained by
prior downstream back-pressure. Section~IV\,A characterizes the minimum
scheduling margin, late-arrival response, multichannel consistency, and
back-pressure behavior.

Absolute and stage-relative timing use the same committed-event path.
Time-oriented instructions update the active temporal reference in the
control core, which resolves the effective timestamp before descriptor
commitment. TDPC therefore uses the same dispatch rule for both timing
modes. Optional per-channel compensation registers can shift the
registered issue cycle by a programmed constant to account for calibrated
fixed latency in downstream electronics.

\subsection{Runtime Adaptation through the Dynamic Overwrite Mechanism}

DOM comprises an Instruction Update Module (IUM), which accepts and
routes runtime-update transactions, and an Instruction Coordination
Module (ICM), which tracks fetch and protects revised program-memory
regions until their contents become safely visible. Protection is local
to instruction fetch.

DOM provides parameter revision through DIPU and structural revision
through DISO.

DIPU affects an event only if the patch is available at parameter capture.
A patch that misses the intended event remains pending for the next
matching event. DISO admission is governed by the hardware safety checks described below. A rejected request raises the update-error status
and preserves the resident instruction path; an admitted request proceeds
in the background or under local fetch protection.

\subsubsection{Dynamic Instruction Parameter Update}

DIPU revises CSR-bound parameters of future peripheral events without
modifying program memory or redirecting instruction fetch. Each update
arrives as a single 64-bit word at the processor-facing
Advanced eXtensible Interface (AXI)-Stream runtime-update interface: bits \texttt{[63:32]} specify the target CSR
address, and bits \texttt{[31:0]} carry the patch value. Acceptance occurs
only when the valid--ready handshake completes. The IUM decodes the
accepted word and forwards the patch transaction to the control core
through the internal \texttt{s\_axis\_csr} interface. Accepted transactions
are serialized in handshake order.

For each patchable CSR entry, the control core maintains a
software-controlled base value, \texttt{base\_csr}, a runtime patch
value, \texttt{patch\_csr}, and a validity state,
\texttt{patch\_valid}. An accepted patch writes
\texttt{patch\_csr} and asserts \texttt{patch\_valid}, while
program-issued CSR writes update only \texttt{base\_csr}. A second
update targeting the same CSR is back-pressured while the patch is
pending; updates to other patchable entries remain independent.

When a peripheral-oriented instruction references a patchable entry,
the event-formation path selects \texttt{patch\_csr} if
\texttt{patch\_valid} is asserted and otherwise selects
\texttt{base\_csr}. Capture of the selected value into the descriptor
defines the parameter-capture boundary. If a patch is selected, it
remains reserved for that descriptor until commitment; during this
interval, no later descriptor referencing the same CSR is formed and
no second same-target patch is accepted. Commitment clears
\texttt{patch\_valid}. Each accepted patch therefore has one-shot,
next-match semantics, while unrelated descriptors and parameter
updates may proceed independently.

DIPU timing is characterized from the processor-facing valid--ready
handshake to parameter capture by the target event. This interval
includes IUM forwarding, clock-domain transfer, patch-state update, and
parameter selection. Section~IV\,B reports the acceptance-to-capture
boundary.

\subsubsection{Dynamic Instruction Stream Overwrite}
\label{sec:diso_architecture}

A DISO transaction begins with an accepted control header followed,
when required, by an operation-dependent sequence of 64-bit
instruction words. The header contains
\texttt{target\_pmem\_addr[31:0]},
\texttt{packet\_type[7:0]}, and
\texttt{length[7:0]}; the remaining bits are reserved. The
implemented packet types are \texttt{8'h00} for contiguous
replacement, \texttt{8'h01} for logical deletion, and
\texttt{8'h02} for out-of-line insertion. For replacement and
deletion, \texttt{length} specifies the number $L$ of consecutive
resident words affected by the transaction. For insertion, it
specifies the number $N$ of inserted instructions.

The supported transaction domain comprises recognized packet types,
lengths of $1\leq L\leq255$ or $1\leq N\leq255$, as applicable, and target
and extension intervals within their configured program-memory (PMEM) regions. Any required
payload must eventually be delivered in full after header acceptance.
Section~\ref{sec:diso_characterization} specifies the subset covered by hardware
measurements. Behavior outside the supported domain is not characterized.

The ICM protects unfinished content from fetch. Timeout, rollback, and
complete-packet buffering are not implemented.

For contiguous replacement, let $A_0$ and $A_1$ denote the first and last target instruction addresses. The target interval is

\begin{equation}
[A_0,A_1], \qquad A_1=A_0+L-1,
\label{eq:diso_target_interval}
\end{equation}

and the $L$ received instruction words replace the corresponding
resident words in address order. Logical deletion uses the same
guarded write path but installs \texttt{nop} encodings over the
selected interval. It therefore suppresses the architectural effects
of the target block without compacting or relocating the remainder
of the resident program.

Logical insertion is implemented through a preallocated out-of-line
extension region beginning at address $B_0$. To insert $N$
instructions before the resident instruction at $A_0$, the extension
region contains the following words, where $I_{\mathrm{old}}(A_0)$ denotes the resident instruction at $A_0$ immediately before this insertion:

\begin{align}
B_0,\ldots,B_0+N-1 &: \text{inserted instructions}, \nonumber\\
B_0+N &: I_{\mathrm{old}}(A_0), \nonumber\\
B_0+N+1 &: \operatorname{jump}(A_0+1).
\label{eq:diso_extension_layout}
\end{align}
The extension block therefore spans
\([B_0,B_{\mathrm{end}}]\), where
\(B_{\mathrm{end}}=B_0+N+1\).

The resident word at $A_0$ is replaced by an unconditional jump to
$B_0$, giving the logical execution path

\begin{equation}
A_0
\rightarrow
[B_0,\ldots,B_0+N-1]
\rightarrow
I_{\mathrm{old}}(A_0)
\rightarrow
A_0+1.
\label{eq:diso_insertion_path_results}
\end{equation}

Out-of-line insertion is permitted only when the resident instruction at
\(A_0\) is relocation-safe and completes by falling through. Its semantics
must not depend on its physical PMEM address, and normal execution must
continue to the next extension word, which contains the return jump.
Program-counter (PC)-relative instructions and instructions that redirect control before
fall-through require compiler-side transformation or are ineligible as
insertion entries. An insertion transaction produces $N+3$ program-memory
writes: the resident entry redirection, the $N$ inserted instructions,
the preserved resident instruction, and the return jump. Subsequent
resident instructions are not shifted. The extension base $B_0$ and the extension-region capacity are
statically configured. The current implementation supports
a single out-of-line insertion for a given feedback response; repeated
insertion using the same extension region is not supported within that
response.

The IUM writes accepted payload words to PMEM in stream order as soon as
they are available, without waiting for the complete payload. If writing
catches up with payload reception, it pauses until the next word is
accepted, then resumes. Under continuous payload delivery, the runtime
write path sustains one 64-bit instruction word per update-domain cycle
after write startup.

Before the first runtime program-memory write, the ICM evaluates the initial fetch distance

\begin{equation}
D_{\mathrm{PC}} = A_0-\mathrm{pc}_{\mathrm{cur}},
\label{eq:diso_pc_distance}
\end{equation}

where $\mathrm{pc}_{\mathrm{cur}}$ is the fetch-side program counter
sampled in the guard-decision cycle. For insertion, $A_0$ denotes the
resident entry instruction that will be redirected to the extension
region. With the configured admission threshold
$G_{\mathrm{guard}}$ and background-start threshold
$G_{\mathrm{bg}}$, the fetch-distance component of the initial mode selection is

\begin{equation}
\operatorname{mode}(D_{\mathrm{PC}})=
\begin{cases}
\mathrm{R}, & D_{\mathrm{PC}}<G_{\mathrm{guard}},\\
\mathrm{S}, &
G_{\mathrm{guard}}\leq D_{\mathrm{PC}}<G_{\mathrm{bg}},\\
\mathrm{B}, & D_{\mathrm{PC}}\geq G_{\mathrm{bg}}.
\end{cases}
\label{eq:diso_admission_policy}
\end{equation}

Here, $\mathrm{R}$ denotes rejection, $\mathrm{S}$ denotes
admission with an immediate fetch stall, and $\mathrm{B}$ denotes
admission as a background write. This distance-based classification is part of
the hardware safety decision, which also checks whether target instructions
have already entered the control-core pipeline.

A rejected request produces no runtime program-memory write, asserts
the update-error status, and leaves the resident execution path
unchanged for the current program instance. The admission threshold
provides a conservative exclusion interval because
$\mathrm{pc}_{\mathrm{cur}}$ is a fetch position rather than a
retirement boundary. The background-start threshold selects only the initial operating mode
of an admitted transaction; it does not replace the continuing visibility
guard.

After admission, the ICM continuously compares the actual fetch
address with the active protected region. This comparison applies to
both sequential execution and branch- or jump-directed fetches.
Consequently, a transaction that begins as a background write may
later transition to stall-protected execution if fetch approaches
revised content that is not yet safely visible.

For contiguous replacement or logical deletion, let
$F_{\mathrm{vis}}(t)$ denote the highest target address whose revised
instruction has completed its program-memory write and is visible to
instruction fetch. Let $t_{\mathrm{w}}$ denote completion of the
required revised-instruction writes, distinct from the transaction-completion
endpoint used in Sec.~IV~C. At time $t$ after activation of the
transaction's protection, let $A_{\mathrm{p}}(t)$ denote the lower protected address. The protected interval $\mathcal{P}(t)$ is

\begin{equation}
\begin{aligned}
A_{\mathrm{p}}(t) &=
\max\!\left(A_0,F_{\mathrm{vis}}(t)-G_{\mathrm{stall}}\right),\\
\mathcal{P}(t) &=
\begin{cases}
[A_{\mathrm{p}}(t),A_1], & t<t_{\mathrm{w}},\\
\varnothing, & t\geq t_{\mathrm{w}}.
\end{cases}
\end{aligned}
\label{eq:diso_protected_interval}
\end{equation}

where $G_{\mathrm{stall}}=5$ instruction words is the configured
local fetch-protection margin. For an accepted DISO transaction,
this margin is used by the ICM to determine whether the
control-core fetch path must be locally stalled as execution
approaches revised instructions that are not yet safely visible.
 Before the first revised
word becomes visible, the mathematical convention
\mbox{$F_{\mathrm{vis}}(t)=A_0-1$} makes the complete target interval protected.
As the write frontier advances, the protected interval moves with it.
When all required revised-instruction writes are complete, the protection
is removed; no protected tail remains after $t_{\mathrm{w}}$.

An insertion transaction protects both the resident entry address
$A_0$ and the configured extension region before writing begins. The same frontier rule is applied to the extension
block by using $B_0$ and $B_{\mathrm{end}}$ as its lower and upper
bounds, respectively; the redirect entry at $A_0$ is protected
separately until it becomes fetch-visible. The
IUM first installs the entry redirection and then writes the
extension block sequentially. The extension target remains
independently protected: if redirected fetch reaches an extension
instruction that is not yet safely visible, the ICM asserts
\texttt{pc\_stall} while program-memory writing continues. Fetch
resumes after the required revised instruction becomes visible.
The same completion rule applies to insertion: no protected interval
remains once all writes to the resident entry and extension block are
complete.

Program-memory writing may continue during a local fetch stall, while
the free-running hardware counter, committed-event FIFOs, and TDPC
dispatch path continue to service events that were committed before
or during the revision. Section~IV\,C characterizes DISO operation.

\subsection{Program Compilation and Runtime-Update Contract}
\label{sec:program_runtime_contract}

A prototype reference assembler translates TIDE assembly into 64-bit
machine-code words for host-side PMEM loading before execution. It
resolves labels and checks supported instruction encodings and operand
ranges. Prototype reference DISO tools construct update transactions and
check operation-specific payload, address, and insertion constraints.

In a prospective feedback path, a decoder would supply the revision operation, target, and operands to an instruction generator. A packetizer would then send the encoded words through the existing DISO header and payload interfaces. The following compatibility rules apply to both generated payloads and precompiled templates.

The checks cover operand and address ranges, operation-specific
length limits, and supported control-flow restrictions. All
position-dependent branch and jump targets in a DISO payload must
resolve within that payload; references to external resident-program
symbols are not supported. TIDE direct branches and JAL use
instruction-word-relative displacements, so references internal to a
payload retain their offsets when the block is relocated. For
insertion, the selected resident entry must additionally satisfy the
relocation-safe and fall-through requirements in Sec.~II~C because the
resident instruction at that entry is relocated to the extension
region. The reference software evaluates insertion-site eligibility
against the instruction currently occupying the selected resident
address, including any earlier revision, together with the configured
address and extension-region constraints. A hardware-generation path would require equivalent
target-eligibility checks before an insertion request is issued.
Encoding constraints can otherwise be enforced by construction, with
runtime checks applied to decision-dependent fields.

The reference tools check instruction and transaction legality; hardware
enforces the runtime boundaries. The control core maintains DIPU patch
state and controls parameter capture; it consumes each selected patch at
event commitment. The ICM enforces DISO admission and instruction-visibility
protection against the current execution state. External software schedules updates with sufficient lead time for intended
operations and handles missed or rejected revisions.

\section{FPGA Implementation and Processor Configuration}
\label{sec:fpga_implementation}

The TIDE processor was implemented on a Xilinx ZCU102
development platform using a Zynq UltraScale+
\texttt{xczu9eg-ffvb1156-2-e} device. The design was synthesized,
placed, and routed with Xilinx Vivado 2022.2. A processing-system interface using the Python-based PYNQ framework loads the resident program
before execution, initializes base parameter state, and provides run
control. Section~IV describes the fabric-resident traffic-generation and
observation logic used for characterization. This logic is outside the
processor contract.

The implementation uses three clock domains. The programmable
control core and ICM operate at 250~MHz, while the TDPC, IUM, and 48-bit
\texttt{time\_abs} counter operate in a 425~MHz timing/update domain. A
100~MHz AXI domain connects the programmable logic to the
processing-system interface. The 250~MHz and 425~MHz clocks are
generated by separate outputs of the same mixed-mode clock manager
(MMCM). The 425~MHz clock has a nominal period of 2.353~ns, which sets the
resolution of the internal hardware time base.
Multi-bit clock-domain transfers use asynchronous FIFOs or explicit
request--acknowledge handshakes, whereas single-bit control and
status paths use synchronizer chains. Table~\ref{tab:processor_config}
summarizes the implemented processor configuration and post-route
timing results.

\begin{table*}[t]
\caption{Implemented configuration of the TIDE FPGA control processor.}
\label{tab:processor_config}
\centering
\begingroup
\small
\setlength{\tabcolsep}{4pt}
\renewcommand{\arraystretch}{1.15}
\begin{tabular}{%
@{}
p{0.21\textwidth}
p{0.73\textwidth}
@{}
}
\toprule
\textbf{Item} & \textbf{Implemented configuration} \\
\midrule

Platform and tools
&
Xilinx ZCU102 (\texttt{xczu9eg-ffvb1156-2-e});
Vivado 2022.2.
\\

Clock domains and time base
&
250~MHz core, 425~MHz timing/update, and 100~MHz AXI;
48-bit \texttt{time\_abs} (2.353~ns resolution).
\\

Control core
&
Five-stage, cache-free, in-order pipeline;
32$\times$32-bit general-purpose registers (GPRs), 48$\times$32-bit CSRs,
64-bit fixed-length instruction set architecture (ISA).
\\

Memory
&
$1024\times64$-bit dual-port PMEM and
$1024\times32$-bit data memory (DMEM);
dedicated fetch and runtime-write PMEM ports.
\\

TDPC
&
2 trigger, 4 data, and 4 wave channels;
32-entry FIFO per channel;
32-, 32-, and 160-bit payloads, respectively.
\\

DIPU interface
&
Independent 64-bit AXI-Stream;
single-word \{CSR address, patch value\} transaction;
one pending patch per target CSR.
\\

DISO interface
&
Independent 64-bit AXI-Stream;
one header followed by 64-bit instruction words;
8-bit length field, with insertion bounded by the configured extension-region capacity; one PMEM write per 425~MHz cycle.
\\

DISO policy
&
$G_{\mathrm{guard}}=5$,
$G_{\mathrm{bg}}=10$,
$G_{\mathrm{stall}}=5$ instruction words.
\\

Post-route timing
&
Setup worst negative slack (WNS): 0.017~ns (425~MHz),
0.067~ns (250~MHz), and 5.296~ns (100~MHz);
Total negative slack (TNS) = 0~ns.
\\

\bottomrule
\end{tabular}
\endgroup
\end{table*}

Dual-port BRAM provides separate 64-bit instruction-fetch and
write ports. The write port accepts host loading only while the control
core is inactive; during execution, only admitted DISO transactions from
the IUM may write PMEM. Data memory, CSR state, and per-channel
committed-event FIFOs are also implemented in programmable logic, with
the capacities in Table~\ref{tab:processor_config}.

TIDE presents registered trigger, data-control, and waveform-parameter
transactions through AXI-Stream interfaces to downstream peripheral-control
or waveform-generation electronics. On-chip capture logic recorded timing
and payloads at these interfaces (Sec.~IV). The results characterize
interface-cycle determinism and payload integrity, excluding electrical
edge jitter, package- or board-level inter-channel skew, and downstream
latency. Section~IV\,A defines the output-handshake condition and
dispatch cycle.

Table~\ref{tab:resource_utilization} reports post-route resources for the
TIDE processor and runtime-control infrastructure, excluding
characterization-only traffic-generation/capture logic and the Zynq
processing-system hard intellectual-property (IP) blocks. In this configuration, \texttt{mul} uses
lookup-table (LUT)-based logic and no digital signal processing (DSP) resources.

\begin{table}[t]
\caption{Post-route programmable-logic resource utilization of TIDE on
the Xilinx ZCU102. The totals include the control core, memories, DOM,
TDPC, fabric-side AXI, clock, reset, and runtime-control infrastructure,
and exclude processing-system hard resources and
characterization-only traffic-generation and capture logic. LUT denotes lookup table; LUTRAM, LUT-based random-access memory; FF, flip-flop; BRAM, block random-access memory; and DSP, digital signal processing. BUFG denotes a global clock-buffer primitive, and MMCM denotes mixed-mode clock manager.}
\label{tab:resource_utilization}
\centering
\begingroup
\small
\setlength{\tabcolsep}{5pt}
\renewcommand{\arraystretch}{1.12}
\begin{tabular}{
@{}
l
r
r
r
@{}
}
\toprule
\textbf{Resource}
&
\textbf{Used}
&
\textbf{Available}
&
\textbf{Utilization (\%)}
\\
\midrule

LUT       & 40\,933 & 274\,080 & 14.93 \\
LUTRAM    & 1\,270  & 144\,000 & 0.88  \\
FF        & 28\,380 & 548\,160 & 5.18  \\
BRAM tile & 4.5      & 912      & 0.49  \\
DSP       & 0       & 2\,520   & 0.00  \\
BUFG      & 5       & 404      & 1.24  \\
MMCM      & 1       & 4        & 25.00 \\
\bottomrule

\end{tabular}
\endgroup
\end{table}

We performed separate post-route sweeps of TDPC channel count and PMEM depth. The per-channel FIFO depth remained at 32 entries, the patchable CSR count at 48, and DMEM at 1024 words. The channel sweep retained the 2:4:4 ratio of trigger, data, and wave channels.

Table~\ref{tab:scaling} reports scaling from 10 to 80 channels and from
1024 to 8192 logical 64-bit PMEM words. All target clocks, including the
425-MHz timing/update clock, met timing. Because the patchable CSR count
was fixed, the channel sweep does not establish proportional scaling of
runtime parameter contexts or update throughput.

The 10-channel/1K-PMEM reference in Table~\ref{tab:scaling} uses the parameterized observation wrapper of the scaling configurations; Table~\ref{tab:resource_utilization} reports the original board-level implementation. The small differences in LUT utilization and WNS reflect these implementation boundaries, not a change in the TIDE configuration.

\begin{table}[t]
\caption{Post-route scaling with TDPC channel count and PMEM depth. BRAM is reported in 36-Kb-equivalent tiles. WNS is the worst setup slack across all constrained clock groups; TNS is zero for every configuration.}
\label{tab:scaling}
\centering
\small
\begin{tabular}{lccc}
\toprule
Configuration & LUT (\%) & BRAM & WNS (ns) \\
\midrule
10 channels, 1K PMEM & 14.94 & 4.5 & 0.023 \\
20 channels, 1K PMEM & 15.33 & 4.5 & 0.023 \\
40 channels, 1K PMEM & 16.11 & 4.5 & 0.001 \\
80 channels, 1K PMEM & 17.59 & 4.5 & 0.033 \\
\midrule
10 channels, 2K PMEM & 14.98 & 6.5 & 0.028 \\
10 channels, 4K PMEM & 14.96 & 9.5 & 0.001 \\
10 channels, 8K PMEM & 14.94 & 16.5 & 0.010 \\
\bottomrule
\end{tabular}
\end{table}

Table~\ref{tab:resource_utilization} reports TIDE's processor and runtime-control hierarchy. Published QICK resource counts include waveform-generation and readout logic~\cite{DiGuglielmo2025QICKHLS}, whereas the cited ARTIQ Zynq report gives event-submission metrics without corresponding LUT, flip-flop, BRAM, and DSP totals~\cite{Lam2021ARTIQZynq}. These reports provide implementation context; their different functional boundaries do not support a normalized processor-area ranking.

\section{Hardware Characterization}
\label{sec:hardware_characterization}

Post-route resource utilization and static timing were obtained from the
Vivado 2022.2 implementation reports. Functional measurements were performed
with the placed-and-routed TIDE design running on a Xilinx ZCU102. A
fabric-resident traffic generator, configured as a decoder emulator, supplied
cycle-controlled DIPU and DISO transactions, while an on-chip integrated logic
analyzer (ILA) and associated observational logic recorded update
acceptance, parameter capture, event commitment, PMEM writes, fetch
activity, local stalls, completion status, and registered output
transactions. These same-FPGA characterization blocks lie outside the TIDE hierarchy
and are excluded from Table~\ref{tab:resource_utilization}. No physical qubits
or external quantum platform were involved.

Unless stated otherwise, cycle indices refer to the 425~MHz timing/update domain,
and acceptance, commitment, and dispatch are defined by valid--ready
handshakes. External acquisition, decision computation, and waveform
generation are outside the measurement scope.

\begin{figure}[!t]
    \centering
    \includegraphics[width=\columnwidth]{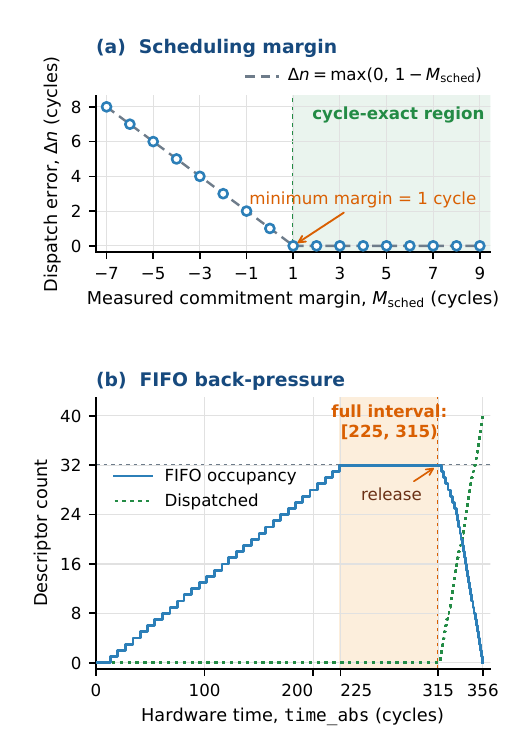}
\caption{Measured committed-event timing and FIFO back-pressure.
(a) Dispatch error $\Delta n$ versus event-commitment margin $M_{\mathrm{sched}}$, both in timing-domain cycles. Fifteen repetitions
were recorded at each of 17 margin settings; coincident records overlap.
All 255 records follow
\(\Delta n=\max(0,1-M_{\mathrm{sched}})\), with programmed-cycle
dispatch for \(M_{\mathrm{sched}}\geq1\).
(b) Occupancy and cumulative dispatch for 40 descriptors traversing a
32-entry TDPC FIFO. The first 32 descriptors fill the FIFO. The remaining eight are held
upstream while the FIFO stays full over
\(225\leq\texttt{time\_abs}<315\). After release, the remaining eight
descriptors commit as space becomes available, and all 40 descriptors
dispatch in order without loss, duplication, reordering, or payload
error.}
    \label{fig:execution_contracts}
\end{figure}

\subsection{Committed-Event Timing and Invariance}
\label{sec:committed_event_characterization}

\subsubsection{Scheduling Margin and Cycle-Exact Dispatch}

The event-commitment cycle
$n_{\mathrm{commit}}$ is defined by successful valid--ready
acceptance of the complete descriptor by the selected TDPC FIFO.
The dispatch cycle $n_{\mathrm{dispatch}}$ is defined by the
corresponding valid--ready handshake at the registered output
interface. Both are recorded relative to the 425~MHz
\texttt{time\_abs} counter. For an event with programmed timestamp
$n_{\mathrm{timestamp}}$, the dispatch error and scheduling margin
are

\begin{equation}
\Delta n
=
n_{\mathrm{dispatch}}-n_{\mathrm{timestamp}},
\label{eq:dispatch_error}
\end{equation}

\begin{equation}
M_{\mathrm{sched}}
=
n_{\mathrm{timestamp}}-n_{\mathrm{commit}}.
\label{eq:scheduling_margin}
\end{equation}

The resident test program generates one trigger descriptor with a fixed
payload and varies its programmed timestamp relative to the measured
commitment cycle. Seventeen settings from $M_{\mathrm{sched}}=-7$ to 9
were evaluated, with 15 repetitions per setting, giving 255 records. The
selected output was held ready; the single-descriptor setup contained neither
an equal-timestamp peer nor an older head-of-line descriptor. At commitment,
the observation logic recorded the timestamp and FIFO payload; at dispatch,
it recorded the output cycle and emitted payload.

Figure~\ref{fig:execution_contracts}(a) shows a single transition
between late and cycle-exact dispatch. All 255 records satisfy

\begin{equation}
\Delta n
=
\max\!\left(0,1-M_{\mathrm{sched}}\right).
\label{eq:measured_dispatch_relation}
\end{equation}

For $M_{\mathrm{sched}}\geq1$, the descriptor is committed at least
one 425~MHz cycle before its timestamp and is dispatched in the
programmed cycle, giving $\Delta n=0$. The minimum measured
commitment margin for cycle-exact dispatch is therefore one
timing-domain cycle, corresponding to 2.353~ns. A descriptor
committed at or after its programmed timestamp is retained and
presented at the first eligible TDPC output cycle rather than being
discarded. Every committed descriptor produced exactly one output
handshake with the committed payload; no payload error, event loss,
or duplicate output occurred in the 255-record data set.

\subsubsection{Post-Commit Invariance}

To test whether later processor activity could alter a committed
descriptor, the resident program formed a target event $E_{0}$ with
timestamp 220 and payload \texttt{0xA5A55A5A}. After $E_{0}$ committed
to the trigger FIFO, one of six perturbations was applied:
branch/jump redirection, interrupt handling, an explicit control-core
stall, a program-issued rewrite of the source CSR, a processor-facing
DIPU transaction, or a DISO instruction revision.

Each perturbation was tested at two placements: shortly after
commitment and near the programmed output cycle. Fifteen repetitions
per perturbation and placement gave
\(6 \times 2 \times 15 = 180\) cases. Records of the six perturbations, including the DIPU acceptance handshake,
confirmed that each occurred after $E_{0}$ commitment.

All 180 target events were dispatched exactly once in timing-domain
cycle 220 with payload \texttt{0xA5A55A5A}. Thus, the target-event dispatch error was \(\Delta n_{0}=0\) in every case, with no timing mismatch, payload
mismatch, event loss, or duplicate output. The free-running
\texttt{time\_abs} counter continued to advance during interrupt
handling, explicit control-core stalls, and DISO activity.

\subsubsection{Multichannel Consistency and FIFO Back-Pressure}

Multichannel consistency was evaluated through the complete
control-core-to-TDPC path. The implemented configuration contains
two trigger, four data-control, and four waveform-parameter
channels. Each channel has an independent 32-entry committed-event
FIFO and a registered AXI-Stream output interface, while all ten
interfaces share the same 425~MHz hardware time base.

Two traffic conditions were characterized. In the idle condition,
each channel received one target event with timestamp 500 and a
channel-specific payload. In the loaded condition, each channel
first received one background event with timestamp 350 and then the
target event at timestamp 500. Each condition was repeated 17 times,
for a total of 34 trials. All 340 target-event handshakes occurred
in timing-domain cycle 500, and all 170 background-event handshakes
occurred in cycle 350. The maximum observed logical
interface-cycle offset among the ten registered interfaces was
therefore zero. All 510 expected transactions were observed with
correct payloads and without missing, duplicate, unexpected, or
cross-channel transactions.

FIFO back-pressure was characterized separately on trigger channel
0 by attempting to commit 40 descriptors with ordered timestamps
and unique payloads through its 32-entry TDPC FIFO. The FIFO read
side was held inactive until the FIFO reached capacity. The FIFO remained full over the half-open interval
$[225,315)$, corresponding to 90 timing-domain cycles. During this interval, the hardware time base
continued to advance and the TDPC FIFO stopped accepting additional
descriptors from the upstream clock-domain-crossing path.

Thirty-two descriptors committed before the FIFO became full. The
remaining eight descriptors were retained upstream until space
became available. After release of the read side, the pending
descriptors committed between cycles 317 and 331; output dispatch
began at cycle 317 and completed at cycle 356. Let $N_{\mathrm{generated}}$, $N_{\mathrm{commit}}$, and $N_{\mathrm{dispatch}}$ denote the generated, committed, and dispatched descriptor counts, respectively. All generated
descriptors traversed the complete exercised path,

\begin{equation}
N_{\mathrm{generated}}
=
N_{\mathrm{commit}}
=
N_{\mathrm{dispatch}}
=
40.
\label{eq:fifo_conservation}
\end{equation}

The descriptor order, timestamps, and payloads were preserved, with
no loss, duplication, reordering, or corruption.
Figure~\ref{fig:execution_contracts}(b) shows FIFO occupancy and
recovery under the 40-descriptor load.

\subsection{Bounded Parameter-Update Visibility}
\label{sec:dipu_characterization}

The DIPU experiment determines how early a parameter update must be accepted to affect its intended event.

The resident program initializes a patchable CSR with the
base value \texttt{0xA5A55A5A} and subsequently forms a trigger
descriptor that references that entry. The traffic generator
supplies the distinct patch value \texttt{0x13579BDF} through the
64-bit DIPU valid--ready interface. The measured path includes
transaction acceptance, IUM forwarding, clock-domain transfer,
pending-patch update, parameter selection, descriptor formation,
event commitment, and registered output dispatch.

The processor-facing valid--ready handshake is mapped to the first
250~MHz control-domain edge not preceding its physical acceptance
time, denoted by $n_{\mathrm{accept,250}}$. Parameter capture occurs
when the event-formation path selects the base or patched value and
captures it into the descriptor, at cycle
$n_{\mathrm{capture,250}}$. The measured
acceptance-to-capture margin is

\begin{equation}
M_{\mathrm{DIPU}}
=
n_{\mathrm{capture,250}}
-
n_{\mathrm{accept,250}}.
\label{eq:dipu_margin}
\end{equation}

The margin includes IUM forwarding and the update-to-control
clock-domain transfer and is expressed in fixed 250~MHz control-domain
cycles; larger values correspond to earlier acceptance.

Seventeen integer-cycle margins from
$M_{\mathrm{DIPU}}=-4$ to 12 were tested, with fifteen repetitions
at each margin, giving 255 transactions. All 120 transactions with
$M_{\mathrm{DIPU}}\leq3$ retained the base value, whereas all
135 transactions with $M_{\mathrm{DIPU}}\geq4$ selected the patch.
The measured boundary
is therefore

\begin{equation}
M_{\mathrm{DIPU,miss}}=3,
\qquad
M_{\mathrm{DIPU,safe}}=4.
\label{eq:dipu_boundary}
\end{equation}

Let $t_{\mathrm{capture}}$ and $t_{\mathrm{accept}}$ denote the physical parameter-capture and update-acceptance times, respectively. Under this edge-mapping convention,
\[
16~\mathrm{ns}
\leq
t_{\mathrm{capture}}-t_{\mathrm{accept}}
<
20~\mathrm{ns}
\]
for \(M_{\mathrm{DIPU}}=4\), depending on the cross-domain phase. The
four-cycle value is the minimum all-success mapped margin over the cross-domain
phases sampled in this clock configuration. A two-event lifecycle test further
confirmed one-shot, next-match semantics: a patch accepted with this margin
was selected by event $E_{0}$, while the following event $E_{1}$, referencing
the same CSR, returned to the base value after the patch was consumed at
$E_{0}$ commitment. No handshake error, payload mismatch, event loss, or
duplicate output occurred across the 255 boundary-characterization
transactions and two lifecycle transactions.

\begin{figure*}[t]
    \centering
    \includegraphics[width=\textwidth]{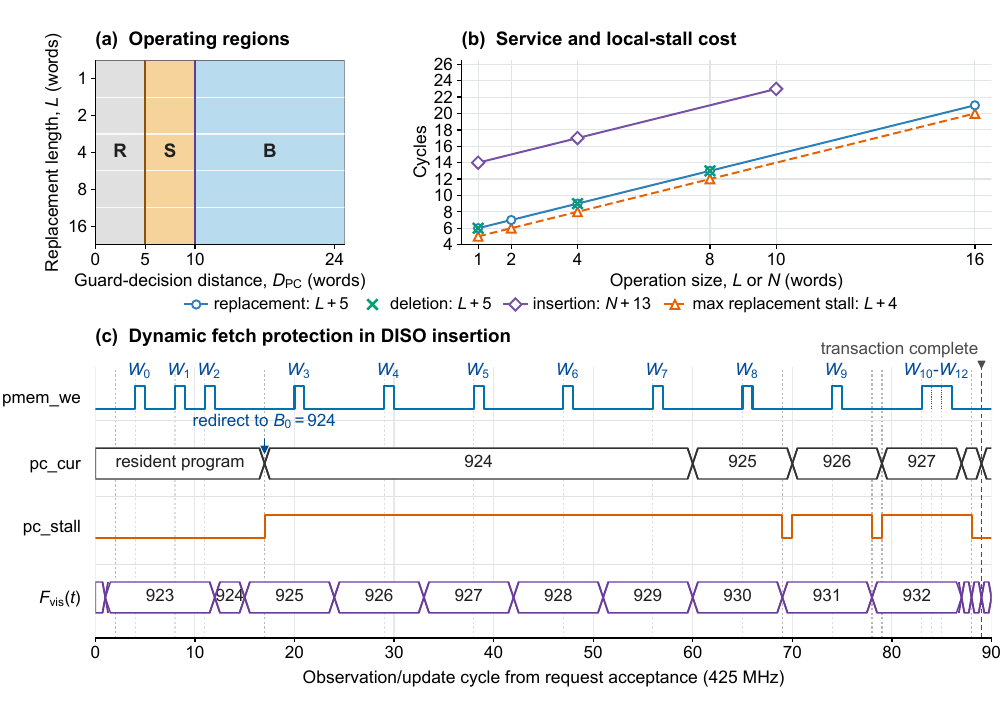}
\caption{Measured guarded runtime instruction revision.
(a) Operating-region summary for 125 replacement cases over
\(L\in\{1,2,4,8,16\}\) and \(D_{\mathrm{PC}}=0,\ldots,24\).
All 125 cases followed the admission rule: rejection for
\(D_{\mathrm{PC}}<5\), immediate local fetch protection for
\(5\leq D_{\mathrm{PC}}<10\), and background start for
\(D_{\mathrm{PC}}\geq10\).
(b) Measured service latency and maximum local-fetch stall under
continuous payload delivery. Replacement latency was \(L+5\) cycles for
\(L\in\{1,2,4,8,16\}\); deletion at \(L\in\{1,4,8\}\) followed the
same law. Insertion latency was \(N+13\) cycles for the tested sizes,
and the maximum observed replacement stall was \(L+4\) cycles.
(c) Gap-injected insertion with \(N=10\) and
\(D_{\mathrm{PC}}=10\). Slowed payload delivery causes an initially
background transaction to enter local fetch protection as redirected
fetch approaches the advancing visibility frontier. Labels W0--W12 identify the 13 program-memory write events.}
    \label{fig:diso_characterization}
\end{figure*}

\subsection{Guarded Runtime Instruction Revision}
\label{sec:diso_characterization}

The DISO measurements characterize runtime structural revision from
processor-facing request acceptance through IUM--ICM coordination,
guarded PMEM modification, control-core fetch, and execution of the
resulting resident or revised path. No host-side PMEM transaction
participates after execution begins. In the characterization below, admission outcomes and initial operating modes
are indexed by the fetch distance sampled at the guard decision,

\begin{equation}
D_{\mathrm{PC}}
=
A_{0}-\mathrm{pc}_{\mathrm{cur}},
\label{eq:measured_pc_distance}
\end{equation}

where $A_{0}$ is the resident target-block entry address and
$\mathrm{pc}_{\mathrm{cur}}$ is sampled in the guard-decision cycle.
Observational logic records request acceptance, PMEM writes,
completion, accumulated \texttt{pc\_stall}, update-error status,
fetch activity, and final architectural signatures. Distinct
resident, revised, and deliberately constructed mixed-sequence
signatures identify the path executed by the control core, while a
continuation marker verifies correct progression beyond the target
block.

\subsubsection{Admission Map and Operating Modes}

Contiguous replacement was characterized first because it exercises
the guarded PMEM-write path shared by the three DISO operations. The
measured matrix covers

\[
L\in\{1,2,4,8,16\},
\qquad
D_{\mathrm{PC}}\in\{0,1,\ldots,24\},
\]

giving 125 block-length--distance combinations. The admission results reported below apply to this tested range. Payload words were
delivered continuously for this
measurement, allowing the runtime write path to sustain one 64-bit
instruction word per 425~MHz update-domain cycle after write
startup.

Figure~\ref{fig:diso_characterization}(a) shows the measured
operating modes. Every tested block length yielded the configured
admission boundaries:

\begin{equation}
\operatorname{outcome}(D_{\mathrm{PC}})
=
\begin{cases}
\mathrm{R}, & D_{\mathrm{PC}}<5,\\
\mathrm{S}, & 5\leq D_{\mathrm{PC}}<10,\\
\mathrm{B}, & D_{\mathrm{PC}}\geq10.
\end{cases}
\label{eq:measured_diso_map}
\end{equation}

Here, $\mathrm{R}$ denotes rejection, $\mathrm{S}$ denotes admission
with immediate local fetch protection, and $\mathrm{B}$ denotes
admission as a background write. No unexpected classification
occurred. Rejected requests asserted the update-error status,
generated no runtime PMEM write, and preserved the resident
execution path. Requests in the intermediate region were admitted
under immediate fetch protection, whereas requests with at least ten
words of initial headroom began as background updates. Under
continuous payload delivery, the visible write frontier remained
ahead of sequential fetch for the background cases, so no subsequent
stall was required.

The two decision boundaries did not change with $L$ over the
measured matrix; block length affected the subsequent service time
and potential stall cost.

\subsubsection{Service Time and Incomplete-Sequence Protection}

For all tested replacement lengths, the measured
request-to-completion latency was

\begin{equation}
N_{\mathrm{overwrite}}(L)
=
L+5.
\label{eq:overwrite_service}
\end{equation}

The length-dependent term corresponds to writing $L$ consecutive
64-bit instruction words at one word per update-domain cycle. The
fixed five-cycle term includes request interpretation, IUM--ICM
coordination, write startup, and PMEM-write visibility. The request-to-completion
latency therefore increased from 6 cycles for $L=1$ to 21 cycles
for $L=16$, corresponding to approximately 14.12--49.41~ns at
425~MHz.

Across the tested contiguous-replacement sweep, the largest accumulated
local fetch stall observed for each block length followed

\begin{equation}
N_{\mathrm{stall,max}}(L)
=
L+4,
\label{eq:maximum_stall_cost}
\end{equation}

over the same tested lengths, increasing from 5 to 20 cycles.
Transactions that remained entirely in the background accumulated
no fetch-stall cycles. Figure~\ref{fig:diso_characterization}(b)
summarizes the measured service and stall costs.

Incomplete-sequence protection was evaluated using distinct
architectural-register signatures for resident, revised, and
deliberately constructed mixed resident/revised execution. Rejected transactions
produced the resident signature, whereas every admitted replacement
produced the complete revised signature. No mixed resident/revised
execution signature was observed. Each admitted transaction also
generated exactly $L$ PMEM writes at the expected addresses.

A separate directed-fetch test exercised the visibility interlock. The control core was held before entering an incompletely
revised interval and resumed only after the required instruction
became safely visible. PMEM could therefore contain both resident
and revised words transiently, but incomplete revised content was
not delivered to instruction fetch.

\subsubsection{Logical Deletion and Out-of-Line Insertion}

Logical deletion was characterized by replacing
$L_{\mathrm{del}}\in\{1,4,8\}$ resident instructions with
\texttt{nop}. Across 204 transactions (68 per length), the 51 cases at
$D_{\mathrm{PC}}=4$ were rejected without a PMEM write, and the remaining
153 were admitted. Every admitted deletion suppressed the target-block
architectural
effects and reached the continuation marker exactly once. Completion
required 6, 9, and 13 update-domain cycles for
$L_{\mathrm{del}}=1$, 4, and 8, respectively, consistent with the
same $L_{\mathrm{del}}+5$ service law measured for contiguous
replacement.

Out-of-line insertion was evaluated for
$N\in\{1,4,10\}$, using the construction in Sec.~\ref{sec:diso_architecture}.
Across 204 insertion transactions (68 per size), the 51 cases at
$D_{\mathrm{PC}}=4$ were rejected without modifying either the resident
entry or the extension region. Every one of the 153 admitted requests
completed the expected
$N+3$ PMEM writes and executed the inserted sequence, preserved
resident instruction, and continuation exactly once and in order.

The measured insertion completion times were 14, 17, and 23
update-domain cycles for $N=1$, 4, and 10, respectively. Over these
tested insertion sizes,

\begin{equation}
N_{\mathrm{insert}}(N)
=
N+13.
\label{eq:insertion_service}
\end{equation}

No missing, duplicated, misordered, or incomplete revised sequence
was observed in the deletion or insertion measurements.

\subsubsection{Dynamic Protection during Throttled Insertion}

The throttled-insertion test examines whether visibility protection remains effective after background admission when payload delivery slows.

An out-of-line insertion with
$N=10$, $D_{\mathrm{PC}}=10$, resident entry
$A_{0}=15$, and extension base $B_{0}=924$ was initially admitted as
a background transaction. Eight idle update-domain cycles were
inserted between successive payload words, slowing the advance of
the visible extension-region write frontier.

As redirected fetch approached instructions that were not yet safely
visible, the ICM asserted \texttt{pc\_stall}; program-memory writing
continued during the stall. As additional extension instructions
became visible, the protected boundary advanced and fetch resumed
intermittently. The insertion generated all 13 expected PMEM writes,
ten of which occurred while \texttt{pc\_stall} was asserted. The
accumulated local fetch stall was 69 cycles. The final
\texttt{pc\_stall} interval ended at relative
cycle 88, and the DISO transaction completed at cycle 89 after all
required writes became visible.

The control core subsequently executed the inserted sequence,
preserved resident instruction, and continuation with the expected
architectural signature. No stale extension instruction,
incomplete revised sequence, or fetch-protection violation occurred.
Figure~\ref{fig:diso_characterization}(c) shows the transition from
background writing to local fetch protection: runtime PMEM writing,
the free-running hardware time base, and dispatch of committed events
continue while only unsafe instruction fetch is held.

\section{Discussion}

TIDE places the effects of runtime adaptation upstream of event commitment. Parameter changes are resolved at capture, and instruction changes are coordinated with the advancing fetch path. Local fetch protection accommodates delayed instruction delivery without stopping the hardware time base or the committed-event service path. The scheduling, post-commitment, and revision tests verify this separation under the exercised execution and update activity.

The two update paths support complementary forms of adaptive quantum control. DIPU binds a one-shot value to the next matching event when available at parameter capture, with persistent settings maintained separately in the base parameter state. DISO supplies or revises instruction blocks in a running program, complementing resident loops and branches. This permits new sequence decisions and calibration-derived values to enter the execution process at the required granularity. DISO admission protects instruction visibility; timely event formation and commitment also require payload delivery and fetch stalls to leave sufficient margin before the programmed timestamps.

Scaling depends on both implementation capacity and update traffic. At a QEC round rate $f_{\mathrm{QEC}}$, $N_{\mathrm{DIPU}}$ single-word parameter updates and $W_{\mathrm{DISO}}$ PMEM word writes per round require $N_{\mathrm{DIPU}}f_{\mathrm{QEC}}$ updates/s and $W_{\mathrm{DISO}}f_{\mathrm{QEC}}$ writes/s, respectively. DIPU and DISO use independent processor-facing interfaces. With continuous payload delivery, DISO writes one 64-bit PMEM word per 425-MHz cycle after startup. An illustrative 1-$\mu$s interval spans 425 update-domain cycles; a continuously supplied 16-word replacement has a measured request-to-completion latency of 21 such cycles (49.41~ns). Transaction overhead and payload gaps reduce useful write throughput. PMEM has separate fetch and runtime-write ports; increasing its depth does not increase their bandwidth.

For DIPU, one accepted word per update-domain cycle gives an ideal ingress ceiling of 425 updates per 1-$\mu$s interval. Actual throughput also depends on forwarding, clock-domain transfer, and patch consumption. The 48 CSR contexts hold simultaneously pending patches, with further updates to a target back-pressured until its selected patch is consumed at commitment. The measured capture margin characterizes update lead time rather than sustained throughput.

The independent capacity sweeps in Table~\ref{tab:scaling} extend the design to 80 channels or 8K PMEM words at the original clock constraints, with FIFO depth, CSR count, and DMEM capacity fixed. Channel allocation and multiplexing determine the mapping from control interfaces to qubits. A prospective multi-core or multi-board design could replicate PMEM, DOM state, and TDPC queues per core, combining local update boundaries with a common time reference~\cite{CarreraVazquez2024RealTimeCommunication} and delivery before local deadlines. Atomic cross-core revision is outside the present implementation. Integration into a quantum experiment requires a feedback budget that also includes acquisition, decoding, communication, event generation, visibility-induced stalls, and downstream waveform generation.

\section{Conclusion}

TIDE makes a running quantum-control program adaptable at two levels: the structure of its future instructions and the numerical parameters of its events. The coordination of these updates with instruction fetch and event commitment preserves already committed descriptors as later control decisions arrive. On ZCU102, the admitted revisions in the characterized cases executed complete updated sequences, while separate post-commit tests verified timestamp and payload invariance. The measured capture margins and revision service costs quantify the processor-side timing requirements for applying updates.

The resulting execution model links runtime decision making to scheduled quantum operations within one control architecture. It lets control programs incorporate both newly supplied operation sequences and event-specific parameter values during execution. This combination provides an execution foundation for measurement-dependent algorithms, adaptive calibration, and the sustained classical feedback required by fault-tolerant quantum computation.

\section*{Acknowledgments}
The authors thank Professor Li You for his support and encouragement, and Zhengran Zhao, Zhicheng Gong, and Tengyu Zhang for helpful discussions. This work was supported by the National Natural Science Foundation of China under Grant No.~92565306.

\section*{Author Declarations}

\subsection*{Conflict of Interest}

The authors have no conflicts to disclose.

\subsection*{Ethics Approval}

Ethics approval was not required for this study.

\subsection*{Author Contributions}

Xiaoqin Luo: Methodology; Validation; Investigation;
Writing -- original draft.

Jiayun Song: Software; Validation; Investigation;
Writing -- original draft.

Xiaolu Su: Conceptualization; Supervision; Project administration;
Writing -- review \& editing.

\section*{Data Availability}
The data that support the findings of this study are available from the corresponding author upon reasonable request.

\bibliography{references}

\end{document}